# Thermal Diffusivity Measurement Based on Thermal / Cooling Excitation: Theory and Experiments

Pengfei Zhu, *Graduate Student Member, IEEE*, Hai Zhang*, *Member*, *IEEE*, Stefano Sfarra, Fabrizio Sarasini, Clemente Ibarra-Castanedo, Xavier Maldague, *Life Senior Member, IEEE*, and Andreas Mandelis

*Abstract*—Photothermal methods for measuring thermal diffusivity inherently pose an ill-posed inverse problem, affected by factors such as sample thickness, heating or cooling time, and excitation energy. Measurement accuracy becomes particularly challenging under non-impulsive pulsed excitation when the observation timescale is comparable to the pulse duration. This is often due to poorly defined pulse shapes, broadened thermal responses, and the absence of clear boundary condition criteria - especially under significant interfacial temperature gradients where natural convection dominates (Grashof number Gr >> 2,000). The classic Parker solution, while widely used, is physically unrealistic as it assumes adiabatic heat flux and shallow-region heat absorption. In this study, we proof the Parker's assumption is equivalent to the Dirac pulse boundary condition in mathematics. Then, we present comprehensive analytical solutions for thermal/cooling responses under Dirac and rectangular pulse excitations. By comparing with eigenfunction-based solutions for well-posed boundary conditions, we show that Parker's solution is only valid prior to the thermal peak. Through dimensionless processing, we further demonstrate that Parker's solution is a limiting case of the rectangular pulse solution as the heating duration approaches zero. Furthermore, we propose a novel excitation approach—evaporative cryocooling—for thermal diffusivity measurement. This method offers a compact, low-cost, and easy-to-implement alternative to conventional excitation schemes. The theoretical model was further validated through comparison with experimental results.

## I. INTRODUCTION

THERMAL diffusivity measurements are critical in the design of instrumentation systems where temperature fluctuations and thermal stress can affect signal responses [1], [2], [3], [4], [5]. Because thermal diffusivity governs the internal heat propagation velocity of materials, local variations in reveal thermophysical inhomogeneities such as impact damage in composites [6]. However, complex failure mechanisms (e.g., fiber fracture, matrix cracking, delamination) and anisotropic material properties significantly hinder accurate diffusivity assessment. Conventional diffusivity measurement techniques, which typically employ lasers or flash lamps as excitation sources, rely on expensive equipment—including infrared detectors and high-power light sources—and must be performed in specialized laboratories. These limitations highlight the urgent need for portable, low-cost, and efficient excitation alternatives for thermal diffusivity characterization.

The thermal diffusivity measurements are commonly classified into frequency-domain and time-domain approaches. Frequency-domain techniques for thermal diffusivity characterization include photothermal emission [7], photothermal beam deflection (PBD) [8], photothermal displacement (PTD) [9], and the thermal-wave cavity (TWC) photopyroelectric method [10]. In photothermal emission method [7], incident modulated radiation heats periodically the sample surface, and an infrared camera records the periodic temperature fluctuations. By comparing the phase shift between original waveform and recorded signals, the thermal diffusivity can be extracted. Similar to the laser flash method, photothermal emission can be affected by heat losses, stray reflections of the incident light, and material transparency to the infrared radiation. Photothermal beam deflection (PBD) [8] relies on the deflection of a probe beam caused by refractive index gradients in the adjacent gas layer, induced by modulated laser heating, to extract the out-of-plane thermal diffusivity of the sample. A simplified working description of the theoretical model used in data fitting of the PBD technique was proposed by Mandelis [11]. The disadvantages of PBD method include environmental sensitivity, roughness of sample surfaces, and a complicated

This work was supported in part by the Natural Sciences and Engineering Research Council of Canada (NSERC) through the CREATE-oN DuTy! Program under Grant 496439-2017, in part by the Discovery Grants Program under Grant RGPIN-2020-04595, in part by the Canada Research Chair in Multi-polar Infrared Vision (MIVIM), in part by the Canada Foundation for Innovation (CFI) Research Chairs Program under Grant 950-230876, and in part by the CFI-JELF program (38794). (Corresponding authors: Andreas Mandelis; Hai Zhang.)

Pengfei Zhu, Hai Zhang, Clemente Ibarra-Castanedo and Xavier Maldague are with the Department of Electrical and Computer Engineering, Computer Vision and Systems Laboratory (CVSL), Laval University, Québec G1V 0A6, Québec city, Canada (e-mail: pengfei.zhu.1@ulaval.ca; hai.zhang.1@ulaval.ca; clemente.ibarra-castanedo@gel.ulaval.ca; xavier.maldague@gel.ulaval.ca). Hai Zhang is also with the Centre for Composite Materials and Structures (CCMS), Harbin Institute of Technology, Harbin 150001, China (hai.zhang@hit.edu.cn).

Stefano Sfarra is with the Department of Industrial and Information Engineering and Economics (DIIIE), University of L'Aquila, I-67100, L'Aquila, Italy. (e-mail: stefano.sfarra@univaq.it).

Fabrizio Sarasini is with the Department of Chemical Engineering Materials Environment & UDR INSTM, Sapienza University of Rome, Rome, Italy. (e-mail: fabrizio.sarasini@uniroma1.it)

Andreas Mandelis is with the Department of Mechanical and Industrial Engineering, Center for Advanced Diffusion-Wave and Photoacoustic Technologies (CADIPT), University of Toronto, Toronto, ON M5S 3G8, Canada, and also with the Department of Mechanical and Industrial Engineering, Institute for Advanced Non-Destructive and Non-Invasive Diagnostic Technologies (IANDIT), University of Toronto, Toronto, ON M5S 3G8, Canada (email: mandelis@mie.utoronto.ca).

calibration process. Photothermal displacement (PTD) [9] measures the surface displacement of a sample caused by periodic heating from a modulated pump laser, with the resulting thermal expansion detected via changes in the reflection angle of an obliquely incident probe beam. Potential error sources include heat exchange between the sample and the surroundings, and misalignment between the pump and probe laser beams [12]. It can be concluded that all these methods require complicated experimental setups, especially with regard to (photo)thermal excitation equipment. The thermal-wave resonant cavity (TWRC; or simply thermal-wave cavity: TWC) method [10] operates by constructing a resonant thermal-wave cavity, where a thin aluminum foil serves as an intensity-modulated laser-induced thermal-wave oscillator, and a pyroelectric polyvinylidene fluoride (PVDF) film functions as both, signal transducer and standing thermal-wave generator. By scanning the modulation frequency, the system exhibits fundamental and higher-order thermal-wave resonances, with overtone amplitudes attenuated due to thermal diffusion characteristics. The TWC technique enables high-precision and accuracy thermal diffusivity measurements. A key advantage of the TWC method is its ability to maintain a fixed modulation frequency while varying the cavity length. This eliminates the need to normalize signals using the system's transfer function, a step required in all frequency-scanning methods.

Time-domain techniques rely on transient responses, with representative examples including the laser flash method [13], transient thermal grating (TTG) [14], and pulsed photothermal displacement techniques [15]. The laser flash technique (Parker's method) [13] is commonly used for thermal diffusivity measurements since it is contactless, non-destructive, and highly accurate [16], [17]. The method uses optical heating as an instantaneous heating source to excite the front surface of a sample, and a thermocouple is employed to record the rear surface temperature variation. The thermal diffusivity can be calculated based on a one-dimensional transmission heat conduction model. The laser flash method requires ultra-short pulse width, uniform heating by the laser, and adiabatically insulated, homogeneous, uniform sample geometry. Additionally, the sample must be opaque to the wavelength of the incident laser beam. To eliminate the radiative and convective heat losses in the laser flash method, Cowan [18] corrected Parker's model by accounting for heat losses. Clark and Taylor [19] proposed a new method for radiation loss correction based solely on the heating portion of the temperature rise curve. However, the measurement uncertainty is also around 5-10% [20]. Transient thermal gratings (TTG) [21] were used to measure the thermal diffusivity in the in-plane direction of a solid sample. The TTG technique uses the interference of two pulsed laser beams to generate a spatially periodic thermal grating on a sample surface, the relaxation dynamics of which is monitored via a probe beam diffraction and is governed by the material's thermal diffusivity. The limitation of this technique is the low efficiency of the diffraction probe beam, which can be overcome by heterodyne detection. However, it is necessary to avoid time delays between the probe and reference beam pulses in heterodyne detection [22], and the experimental uncertainty is around 10-15% [22]. Pulsed photothermal displacement techniques measure the thermal diffusivity of a sample via its surface deformation caused by a heating pulse. A typical pulsed photothermal displacement method is the pulsed photothermal mirror [15]. In this technique, a single pulse heats the sample and causes subsequent deformation due to thermal expansion. As the sample surface becomes deformed, a probe beam is focused or defocused due to the thermal mirror effect [15], [23]. The thermal diffusivity can be calculated by analyzing the change in the intensity profile of the central portion of the beam in the far field. However, the pulsed photothermal mirror technique requires foreknowledge of Poisson's ratio, surface reflectivity, and extremely low roughness leading to a measurement uncertainty of about 15%.

Recent advances have explored the integration of deep learning with photothermal techniques to estimate thermal diffusivity. In one approach, simulated datasets are employed to train neural networks, which are subsequently applied to experimental measurements for prediction [24]. Alternatively, experimental data have been directly used for network training [25]. While these methods demonstrate the potential of data-driven modeling, they often suffer from limited robustness and reduced accuracy when applied to diverse experimental conditions.

In this study, we derive several analytical solutions—including both practical approximations and mathematically rigorous formulations—for the thermal response of a solid subjected to Dirac-pulse and step-function excitations under photothermal/evaporative cryocooling conditions. Although Lei et al. [26] employed cooling excitation to detect subsurface defects, their work lacked a theoretical framework. In contrast, the proposed method offers a low-cost and easy-to-implement alternative to conventional techniques. A comprehensive evaluation was carried out through both simulations and experiments, and the results were benchmarked against the classic laser flash method to validate accuracy and reliability.

## II. Theory

Several heat transfer formalisms are developed under thermal and cooling modes, such as classic heat transfer models (Parker's solution [13] and Almond's solution [27]) and a Green's function solution derived by Mandelis [29]. The first solution was derived from Carslaw and Jaeger's semi-analytical formulation, which has been widely adopted in practical industrial applications. The second approach was based on the eigenvalue method, offering a more mathematically rigorous and complete solution framework. The Mandelis' Green function method is shown in the Supplementary.

### A. Classic Heat Conduction Model – Dirac Pulse Excitation and Rectangular Pulse Excitation

The classic heat conduction models were derived from Carslaw and Jaeger's formalism [28]. For the impulse heating response of a semi-infinite half space, the boundary-value problem (BVP) can be given as:

$$\begin{cases} \frac{\partial T}{\partial t} = \alpha \frac{\partial^2 T}{\partial x^2}, \ x > 0, t > 0 \\ T(x,0) = 0 \\ -k \frac{\partial T}{\partial x}\big|_{x=0} = J_0 \delta(t) \end{cases} \quad (1)$$

This BVP can be directly solved by the Laplace transform method [28]:

$$T(x,t) = \frac{J_0}{\sqrt{\pi \rho c k t}} e^{\frac{-x^2}{4\alpha t}} \quad (2)$$

where $T(x, t)$ is the temperature rise at a depth $x$ beneath the surface at a time $t$ after a uniform impulse of energy $J_0$ on the surface, $x = 0$, at time $t = 0$. $\rho$, $c$, $k$, and $\alpha$ are density, heat capacity, thermal conductivity, and thermal diffusivity, respectively.

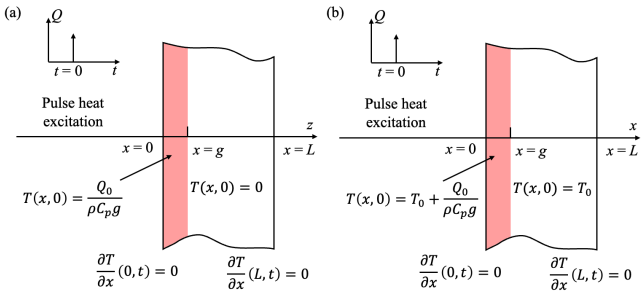

**Fig. 1.** Theoretical model for (a) Parker's solution and (b) Parker's solution with initial temperature. $Q$ denotes impulse energy, $t$ denotes time, $T$ denotes temperature, $T_0$ denotes initial temperature, $g$ denotes the heat absorption depth after pulse excitation, $L$ denotes sample's thickness.

For an adiabatic sample (see Fig. 1(a)) with an initial temperature distribution $f(x)$, the BVP can be given as:

$$\begin{cases} \frac{\partial T}{\partial t} = \alpha \frac{\partial^2 T}{\partial x^2}, 0 < x < L, t > 0 \\ \frac{\partial T}{\partial x}(0,t) = 0, \frac{\partial T}{\partial x}(L,t) = 0 \\ T(x,0) = f(x) \end{cases} \quad (3)$$

where $L$ is the thickness of the specimen. The solution was originally developed by Carslaw and Jaeger in the form of the temperature distribution at time $t$ [28]:

$$T(x,t) = \frac{1}{L}\int_0^L f(x)dz + \frac{2}{L}\sum_{n=0}^{\infty} \exp\left(\frac{-n^2\pi^2\alpha t}{L^2}\right)\cos\left(\frac{n\pi x}{L}\right)\int_0^L \frac{f(x)\cos(n\pi x)}{L}dx \quad (4)$$

In the case of impulse heating, rather than modeling the incident heat flux through a Robin boundary condition, we consider a simplified scenario where rapid heat absorption takes place within a shallow surface layer of the sample. The initial temperature distribution is assumed to be $f(x) = \frac{Q_0}{\rho C_p g}$ for $0 < x < g$, and 0 for $g < x < L$. A pulse of radiant energy $Q(x,t) = Q_0 g \delta(t)$ cal/cm$^2$ is assumed to be instantaneously and uniformly absorbed within a small depth $g = 1/\beta$ of the specimen where $\beta$ is the optical absorption coefficient of the material at the wavelength of the excitation laser, as shown in Fig. 1(b) with the initial temperature set to $T_0$. Therefore, the initial temperature distribution is given as:

$$f(x) = \begin{cases} T_0 + \frac{Q_0}{\rho C_p g}, & 0 < x < g \\ T_0, & g < x < l \end{cases} \quad (5)$$

Substituting Eq. (5) into Eq. (4), the impulse heat response can be reformulated as

$$T(x,t) = T_0 + \frac{Q_0}{\rho C_p L} + 2\frac{Q_0}{\rho C_p g}\sum_{n=1}^{\infty}\frac{1}{n\pi}\cos\left(\frac{n\pi x}{L}\right)\sin\left(\frac{n\pi g}{L}\right)\exp\left(\frac{-n^2\pi^2\alpha t}{L^2}\right) \quad (6)$$

This analytical solution is similar to Parker's solution [13], the only difference being the initial temperature term $T_0$. Therefore, it is consistent with the principle of superposition where the temperature field is a linear combination of the initial temperature $T_0$ and the variation $\Delta T$ caused by the external source. For the transmission mode, i.e., at $x = L$ the simplification $\sin\left(\frac{n\pi g}{L}\right) \approx \frac{n\pi g}{L}$ can be used since $g$ is a very small number for opaque materials and the thermal decay rate is rapid enough to render high-n value terms negligible in the simplified analysis. Therefore, the solution can be reduced to

$$T(L,t) = T_0 + \frac{Q_0}{\rho C_p L}\left[1 + 2\sum_{n=1}^{\infty}(-1)^n \exp\left(\frac{-n^2\pi^2\alpha t}{L^2}\right)\right] \quad (7)$$

This analytical solution may seem to be physically unreasonable as the temperature $T(x, t)$ will tend to $T_0 + \frac{Q_0}{\rho C_p L}$ when $t \to \infty$. Actually, this solution captures the full transient evolution of the temperature distribution following impulse excitation, including the short-time thermal gradients and the eventual approach to thermal equilibrium. First of all, it obeys the energy conservation since the total energy in the initial stage is $T_0 + \frac{Q_0}{\rho C_p L}$. In practical experiments, interfacial heat exchange such as convection and radiation can be neglected on a short time scale, which will be validated in the following part. Eqs. (6) and (7) describe the distribution of heat within the solid until the whole sample reaches thermal equilibrium. However, this solution is not physically realistic because it cannot come back to the ambient temperature at equilibrium. Additionally, this solution is not suitable for the heat source with complex shaped and multi-layered structures and does not consider the effects of emissivity, reflectivity, etc. Therefore, Eq. (7) cannot be used for the determination of the absolute temperature across the sample and on its surface. Nevertheless, the advantages of this analytical solution will be explained in the next Section.

Based on the analytical solution Eq. (7), it is possible to calculate the heat response under rectangular pulse excitation. According to Duhamel's theorem:

$$T(x,t) = \int_0^t G(x,t-\tau)f(x,\tau)d\tau \quad (8)$$

where $G(x, t-\tau)$ is the unit impulse response (Green's function) of the system, $f(x,\tau) = \frac{Q_0}{\rho C_p}\delta(x)H(t)H(\tau-t)$ is the time-dependent heat sink at position x and time $\tau$, and $H(t)$ is the Heaviside unit step function. Generally, the temperature response to long-pulse excitation should be computed using the Green function for a unit impulse. However, to simplify the analysis, we follow the method introduced by Almond et al. [27], wherein the rectangular pulse excitation is convolved directly with the thermal response described in Eq. (7). It should be noted that $T_0$ is not included in the integration, as it represents the system's initial condition, whereas the second term constitutes part of the system's time-dependent response function.

$$T(L,t) = T_0 + \frac{Q_0}{\rho C_p L}[t + 2\sum_{n=1}^{\infty}(-1)^n \frac{L^2}{n^2\pi^2\alpha}(1 - \exp(\frac{-n^2\pi^2\alpha t}{L^2}))] \quad \text{for } t \leq \tau_p \quad (9)$$

$$T(L,t) = T_0 + \frac{Q_0}{\rho C_p L}[\tau_p + 2\sum_{n=1}^{\infty}(-1)^n \frac{L^2}{n^2\pi^2\alpha}(\exp\left(\frac{-n^2\pi^2\alpha(t-\tau_p)}{L^2}\right) - \exp\left(\frac{-n^2\pi^2\alpha t}{L^2}\right))] \quad \text{for } t \geq \tau_p \quad (10)$$

where $\tau_p$ is the pulse duration time. The above equations can be consolidated as follows:

$$T(L,t) = T_0 + \frac{Q_0}{\rho C_p L}[\min(t, \tau_p) + 2\sum_{n=1}^{\infty}(-1)^n \frac{L^2}{n^2\pi^2\alpha}[\exp\left(\frac{-n^2\pi^2\alpha(t-\min(t,\tau_p))}{L^2}\right) - \exp\left(\frac{-n^2\pi^2\alpha t}{L^2}\right)]] \quad (11)$$

*B. Heat Conduction Model Without Considering Convection – Dirac Pulse Excitation and Rectangular Pulse Excitation*

The previous (Parker's) solution assumes adiabatic boundaries and a step-function initial condition in space, corresponding to instantaneous heating at the sample surface. This assumption is widely used in conventional heat transfer models, for example, in problems involving immersion of solids into hot or cold baths modeled as infinite reservoirs, where the boundary temperature is assumed to change instantaneously. It has also been shown to be adequate for very fast thermal inputs without optical excitation, where within the exposure time an infinitesimally thin surface layer experiences rapid (instantaneous) temperature rise or fall, followed by a progressively delayed bulk response. This assumption has been proven to work well and has been found to agree with experimental data in wide ranges of solid geometries discussed in detail in textbooks on heat transfer over many years. Nevertheless, in situations where the heating or cooling duration is comparable to the thermal diffusion time, or when ultrafast volumetric energy deposition occurs (e.g., under optical excitation), or in cases where the pulse duration is commensurate and competes with the effective surface layer rise- or decay-time, such an assumption may underestimate the temporal evolution of the interior thermal field.

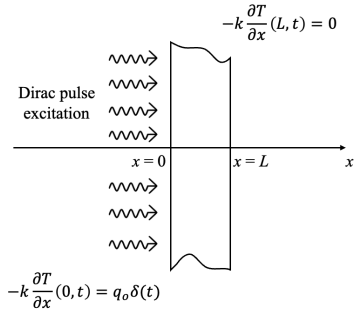

**Fig. 2.** Theoretical model for heat transfer without considering convection. $q_0$ denotes impulse energy, $t$ denotes time, $T$ denotes temperature, $T_0$ denotes initial temperature, $k$ denotes heat conductivity, $L$ denotes sample's thickness.

Therefore, in this work we provide a closer look by assuming a finite-time heat flux into (out of) the front surface and an adiabatic condition on the back surface (without considering convection), as shown in Fig. 2. The boundary-value problem (BVP) can be defined as:

$$\begin{cases} \frac{\partial T}{\partial t} = \alpha \frac{\partial^2 T}{\partial x^2}, \quad x > 0, t > 0 \\ -k\frac{\partial T}{\partial x}|_{x=0} = q_0\delta(t), -k\frac{\partial T}{\partial x}|_{x=L} = 0, \\ T(x,0) = T_0 \end{cases} \quad (12)$$

To simplify the derivation process, let

$$\theta(x,t) = T(x,t) - T_0 \quad (13)$$

The BVP become:

$$\begin{cases} \frac{\partial \theta}{\partial t} = \alpha \frac{\partial^2 \theta}{\partial x^2}, \quad 0 < x < L, t > 0 \\ -k\frac{\partial \theta}{\partial x}|_{x=0} = q_0\delta(t), -k\frac{\partial \theta}{\partial x}|_{x=L} = 0, t > 0 \\ \theta(x,0) = 0 \end{cases} \quad (14)$$

Considering the eigenvalue problem:

$$\phi''(x) + \lambda^2 \phi(x) = 0 \quad (15)$$

With homogeneous Neumann boundary conditions:

$$\phi'(0) = 0, \phi'(L) = 0 \quad (16)$$

The eigenvalue can be given as:

$$\lambda_n = \frac{n\pi}{L}, \phi_n(x) = \cos\left(\frac{n\pi}{L}x\right), \quad \text{for } n = 0,1,2,\ldots \quad (17)$$

where $n = 0$ corresponds the constant eigenfunction $\phi_0(x) = 1$. Then, the temperature response can be given as:

$$\theta(x,t) = \sum_{n=0}^{\infty} A_n(t)\phi_n(x) = A_0(t) + \sum_{n=1}^{\infty} A_n(t)\cos\left(\frac{n\pi}{L}x\right) \quad (18)$$

Substitute Eq. (18) into Eq. (14):

$$\frac{dA_n}{dt} + \alpha\lambda_n^2 A_n = \frac{2}{\rho C_p L}q_0\delta(t)\phi_n(0) \quad (19)$$

For $n \geq 1$ modes:

$$\frac{dA_n}{dt} + \alpha\lambda_n^2 A_n = \frac{2q_0}{\rho C_p L}\delta(t) \quad (20)$$

Using the integrating factor method, we can obtain:

$$A_n(t) = [A_n(0^-) + \frac{2q_0}{\rho C_p L}]e^{-\alpha\lambda_n^2 t} \quad (21)$$

Of note, the temperature equals 0 before initial time, therefore, $A_n(0^-) = 0$ and $A_n(0^+) = \frac{2q_0}{\rho C_p L}$. This gives:

$$A_n(t) = \frac{2q_0}{\rho C_p L}e^{-\alpha\lambda_n^2 t} \quad \text{for } n \geq 1 \quad (22)$$

For $n = 0$ mode, we have:

$$\frac{dA_0}{dt} = \frac{q_0}{\rho C_p L}\delta(t) \Rightarrow A_0(t) = \frac{q_0}{\rho C_p L}, t > 0 \quad (23)$$

Now, we superpose each mode:

$$T(x,t) = T_0 + \frac{q_0}{\rho C_p L} + \sum_{n=1}^{\infty}\frac{2q_0}{\rho C_p L}e^{-\alpha\lambda_n^2 t}\cos\left(\frac{n\pi x}{L}\right) \quad (24)$$

Eq. (24) can be simplified as:

$$T(x,t) = T_0 + \frac{q_0}{\rho C_p L}\left(1 + 2\sum_{n=1}^{\infty}\cos\left(\frac{n\pi x}{L}\right)\exp\left(-\alpha\frac{n^2\pi^2}{L^2}t\right)\right) \quad (25)$$

where $\lambda_n = \frac{n\pi}{L}$ is the eigenvalue. It can be shown that the formalisms in Eq. (13) and Eq. (7) are the same. This means that Parker's assumption is mathematically equivalent to the Dirac pulse heat injection. The proof is given in the Supplement.

## C. Heat Conduction Model Considering Convection – Dirac Pulse Excitation and Rectangular Pulse Excitation

The previous model does not consider convection. The results will become inaccurate if the convective coefficient $h$ or Biot number $Bi$ is high.

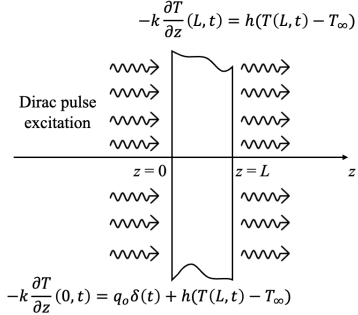

**Fig. 3.** Theoretical model for proper heat conduction model. $q_0$ denotes impulse energy, $t$ denotes time, $T$ denotes temperature, $h$ denotes convective coefficient, $T_\infty$ denotes ambient temperature, $k$ denotes heat conductivity, $L$ denotes sample's thickness.

Considering a more general case for the BVP under impulse excitation (see Fig. 3) with the initial temperature of the sample being the same as the ambient temperature under thermal equilibrium:

$$\begin{cases} \frac{\partial T}{\partial t} = \alpha \frac{\partial^2 T}{\partial x^2}, 0 < x < L, t > 0 \\ -k \frac{\partial T}{\partial x}|_{x=0} = q_0 \delta(t) + h(T(0,t) - T_\infty), \ t > 0 \\ -k \frac{\partial T}{\partial x}|_{x=L} = h(T(L,t) - T_\infty), \ t > 0 \\ T(x,0) = T_\infty, \ 0 \le x \le L \end{cases} \quad (26)$$

Here $h$ is the linearized convective coefficient, $q_0$ is the energy of the Dirac delta-function impulse, and $T_\infty$ is the ambient temperature. We only consider the temperature variation, thus letting $\theta(x,t) = T(x,t) - T_\infty$. The BVP can be reduced to:

$$\begin{cases} \frac{\partial \theta}{\partial t} = \alpha \frac{\partial^2 \theta}{\partial x^2}, \ 0 < x < L, t > 0 \\ -k \frac{d\theta}{dx}|_{x=0} = q_0 \delta(t) + h\theta(0,t), \ t > 0 \\ -k \frac{d\theta}{dx}|_{x=L} = h\theta(L,t), \ t > 0 \\ \theta(x,0) = 0, \ 0 \le x \le L \end{cases} \quad (27)$$

For BVP in Eq. (27), we expand the solution as a modal sum:

$$\theta(x,t) = \sum_{n=1}^{\infty} C_n(t) \phi_n(x) \quad (28)$$

where $\phi_n(x)$ satisfies the eigenvalue problem:

$$\frac{d^2 \phi_n}{dx^2} + \lambda_n^2 \phi_n = 0 \quad (29)$$

Substituting Robin boundary condition on the left side and the right side, yields the general solution:

$$\phi_n(x) = A_n \cos(\lambda_n x) + B_n \sin(\lambda_n x) \quad (30)$$

subject to the eigenvalue equation:

$$\tan(\lambda_n L) = \frac{2(\frac{h}{k})\lambda_n}{\lambda_n^2 - (h/k)^2} \quad (31)$$

The normalization coefficient is:

$$\langle \phi_n, \phi_n \rangle = \int_0^L [A_n \cos(\lambda_n x) + B_n \sin(\lambda_n x)]^2 dx$$
$$= \frac{L}{2} \left(1 + \frac{\left(\frac{h}{k}\right)^2}{\lambda_n^2}\right) + \frac{h}{2k\lambda_n^2}[1 - (-1)^n e^{-2hL/k}] \quad (32)$$

At $t = 0$, the boundary flux pulse satisfies:

$$\int_{0^-}^{0^+} -k \frac{d\theta}{dx}|_{x=0} dt = q_0 \quad (33)$$

Projecting to the $n$-th mode and using orthogonality yields:

$$C_n(0^+) = \frac{q_0}{\rho c} \frac{\phi_n(0)}{\langle \phi_n, \phi_n \rangle} = \frac{2q_0[(k\lambda_n)^2 + h^2]}{\rho c [L((k\lambda_n)^2 + h^2) + hk]} \quad (34)$$

Using Duhamel's theorem, the time-domain evolution of the $n$-th mode is found to be:

$$C_n(t) = C_n(0^+) e^{-\alpha \lambda_n^2 t} \quad (35)$$

Therefore, the final solution is:

$$\theta(x,t) = \sum_{n=1}^{\infty} \left[\frac{2q_0[(k\lambda_n)^2 + h^2]}{\rho c [L((k\lambda_n)^2 + h^2) + hk]}\right] \cos(\lambda_n x) e^{-\alpha \lambda_n^2 t} \quad (36)$$

where $\{\lambda_n\}$ is the positive eigenvalue root satisfying Eq. (31). Likewise, the instantaneous step-function response $(q_0 H(t))$ can be calculated using Duhamel's theorem:

$$\theta_{step}(x,t) =$$
$$\sum_{n=1}^{\infty} \left[\frac{2q_0[(k\lambda_n)^2 + h^2]}{\rho c [L((k\lambda_n)^2 + h^2) + hk]}\right] \cos(\lambda_n x) \int_0^t e^{-\alpha \lambda_n^2 (t-\tau)} d\tau$$
$$= \sum_{n=1}^{\infty} \left[\frac{2q_0[(k\lambda_n)^2 + h^2]}{\rho c [L((k\lambda_n)^2 + h^2) + hk]} \cdot \frac{1 - e^{-\alpha \lambda_n^2 t}}{\alpha \lambda_n^2}\right] \cos(\lambda_n x)$$
$$\text{for } t \le \tau_p \quad (37)$$

$$\theta_{step}(x,t) = \sum_{n=1}^{\infty} \left[\frac{2q_0[(k\lambda_n)^2 + h^2]}{\rho c [L((k\lambda_n)^2 + h^2) + hk]} \cdot \frac{(1 - e^{-\alpha \lambda_n^2 \tau_p}) e^{-\alpha \lambda_n^2 (t - \tau_p)}}{\alpha \lambda_n^2}\right] \cos(\lambda_n x) \text{ for for } t > \tau_p \quad (38)$$

## D. Solution of Cooling Excitation – Dirac Pulse Excitation and Rectangular Pulse Excitation

The physical outward heat transfer model developed for evaporative cryocooling is illustrated in Fig. 4(a). In evaporative cryocooling, surface temperature reduction can occur rapidly due to the latent heat loss during a phase transition. This process can be modeled as a rapid extraction of thermal energy from a shallow subsurface layer, resulting in a sudden local temperature drop. Considering a simplified form – classic heat conduction model (Fig. 4(a)) under surface cooling, the initial condition is modified to:

$$f(x) = \begin{cases} T_0 - \frac{Q_0}{\rho C_p g}, & 0 < x < g \\ T_0, & g < x < l \end{cases} \quad (39)$$

where $T_0$ is the uniform temperature of the material at $t \le 0$. Similarly, for a heat conduction model without convection (Fig. 4(b)) the solution for cooling-induced pulsed excitation of the surface can be expressed as:

$$T(x,t) = T_0 - \frac{Q_0}{\rho C_p L} - 2 \frac{Q_0}{\rho C_p g} \sum_{n=1}^{\infty} \frac{1}{n\pi} \cos\left(\frac{n\pi x}{L}\right) \sin\left(\frac{n\pi g}{L}\right) \exp\left(\frac{-n^2 \pi^2 \alpha t}{L^2}\right) \quad (40)$$

For the heat transmission mode at the back surface, i.e., $x = L$, the solution can be reduced to

$$T(L,t) = T_0 - \frac{Q_0}{\rho C_p L}\left(1 + 2\sum_{n=1}^{\infty}(-1)^n \exp\left(\frac{-n^2\pi^2 \alpha t}{L^2}\right)\right) \quad (41)$$

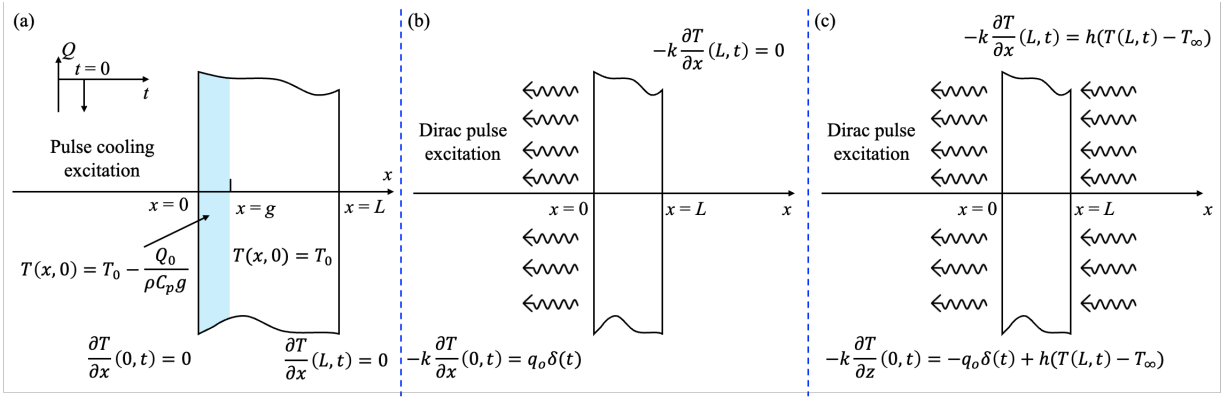

**Fig. 4.** Schematic of (a) classic heat conduction model, (b) heat conduction model without considering convection, and (c) heat conduction model considering convection under evaporative cooling excitation. $q_0$ denotes impulse energy, $t$ denotes time, $T$ denotes temperature, $h$ denotes convective coefficient, $T_\infty$ denotes ambient temperature, $T_0$ denotes initial temperature, $k$ denotes heat conductivity, $L$ denotes sample's thickness.

In the case of instantaneous step-function cooling excitation, the analytical solution can be solved again using Duhamel's theorem:

$$T(L,t) = T_0 - \frac{Q_0}{\rho C_p L}\left[t + 2\sum_{n=1}^{\infty}(-1)^n \frac{L^2}{n^2\pi^2\alpha}\left(1 - \exp\left(\frac{-n^2\pi^2\alpha t}{L^2}\right)\right)\right] \quad \text{for } t \leq \tau_p \quad (42)$$

$$T(L,t) = T_0 - \frac{Q_0}{\rho C_p L}\left[\tau_p + 2\sum_{n=1}^{\infty}(-1)^n \frac{L^2}{n^2\pi^2\alpha}\left(\exp\left(\frac{-n^2\pi^2\alpha(t-\tau_p)}{L^2}\right) - \exp\left(\frac{-n^2\pi^2\alpha t}{L^2}\right)\right)\right] \quad \text{for } t > \tau_p \quad (43)$$

where $\tau_p$ is the pulse duration. The above equations can be written compactly as follows:

$$T(L,t) = T_0 - \frac{Q_0}{\rho C_p L}\left[\min(t, \tau_p) + 2\sum_{n=1}^{\infty}(-1)^n \frac{L^2}{n^2\pi^2\alpha}\left(\exp\left(\frac{-n^2\pi^2\alpha(t-\min(t,\tau_p))}{L^2}\right) - \exp\left(\frac{-n^2\pi^2\alpha t}{L^2}\right)\right)\right] \quad (44)$$

Likewise, considering a physically "proper" form (i.e. with boundary convection and heat input) in Section 2.3, the Dirac impulse response can be calculated (see Fig. 4(c)) as follows:

$$\theta(x,t) = \sum_{n=1}^{\infty}\left[\frac{-2q_0[(k\lambda_n)^2 + h^2]}{\rho c[L((k\lambda_n)^2 + h^2) + hk]}\right]\cos(\lambda_n x)e^{-\alpha\lambda_n^2 t} \quad (45)$$

The instantaneous step-function response can also be shown to be:

$$\theta_{step}(x,t) = \sum_{n=1}^{\infty}\left[\frac{-2q_0[(k\lambda_n)^2 + h^2]}{\rho c[L((k\lambda_n)^2 + h^2) + hk]} \cdot \frac{1 - e^{-\alpha\lambda_n^2 t}}{\alpha\lambda_n^2}\right]\cos(\lambda_n x) \quad \text{for } t \leq \tau_p \quad (46)$$

$$\theta_{step}(x,t) = \sum_{n=1}^{\infty}\left[\frac{-2q_0[(k\lambda_n)^2 + h^2]}{\rho c[L((k\lambda_n)^2 + h^2) + hk]} \cdot \frac{(1 - e^{-\alpha\lambda_n^2\tau_p})e^{-\alpha\lambda_n^2(t-\tau_p)}}{\alpha\lambda_n^2}\right]\cos(\lambda_n x) \quad \text{for } t > \tau_p \quad (47)$$

To validate the accuracy of different solutions, the normalized analytical results from Parker's solution or the heat conduction model without convection (Eqs. (41)-(44)) were compared with the solution considering convection (Eqs. (45)-(47)). The material parameters used are shown in Table 1 which was constructed using general carbon fiber reinforced polymers (CFRP).

TABLE I
The Material Properties for Simulation.

| Material parameter | Value |
|---|---|
| Heat Conductivity (W/(m·K)) | 1.56 |
| Density (kg/m³) | 1800 |
| Pulse energy (J/cm²) | 1000 |
| Convective coefficient (W/(m²·K)) | 50 |
| Material thickness (mm) | 3 |
| Density (kg/m³) | 7800 |
| Heat Capacity (J/(kg·K)) | 500 |
| Thermal Diffusivity (mm²/s) | 4×10⁻⁷ |

It is obvious that the temperature profiles of Parker's solution and the eigenvalue solution agree well before the peak point, which holds for Dirac, step, and cooling excitations. However, for $t \gtrsim 15$ s, the two solutions begin to diverge. This difference originates from the fact that Parker's solution neglects higher-order eigenmodes and assumes an instantaneous boundary temperature change as discussed above, which becomes less accurate when slow heating diffusive effects dominate.

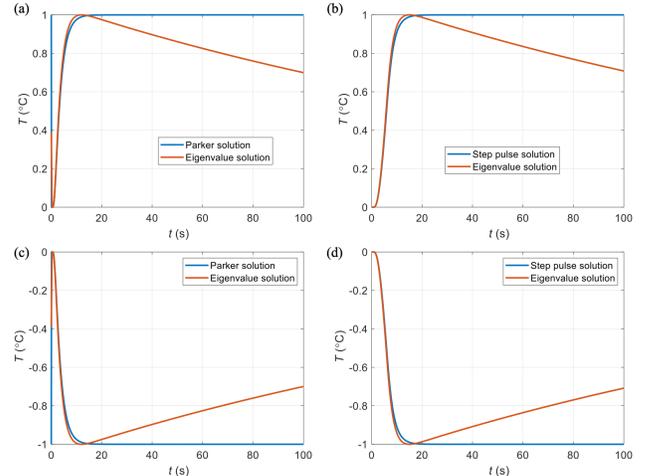

**Fig. 5.** Comparison between Parker's solution and eigenvalue solution at rear surface ($z = L$): (a) Thermal Dirac pulse excitation; (b) instantaneous thermal step-function excitation; (c) Cooling Dirac pulse excitation; (d) Cooling instantaneous step-function excitation.

In the eigenfunction - eigenvalue formulation, the long-time tail reflects gradual heat redistribution within the sample, leading to a slower decay or, in the case of cooling, a slower temperature recovery. In this model, no further heating of the solid is expected at long times once the transient pulse energy has been fully absorbed. Instead, the temperature approaches a uniform steady state. All temperature profiles in Fig. 5 are evaluated at $x = L$. The frequency-domain solution for Dirac pulse thermal/cooling excitation is shown in the Supplement.

## III. THERMAL DIFFUSIVITY MEASUREMENT METHODOLOGIES

### A. Conventional Laser Flash Method

According to the principles of the laser flash method, thermal diffusivity can be determined from the thermal response of the rear face of a sample after the front face is subjected to a laser or flash lamp pulse. Theoretically, the temperature rise on the rear surface as a function of time can be calculated using Eq. (7) which can be reduced to:

$$W(\eta) = 1 + 2 \sum_{n=1}^{\infty} (-1)^n \exp[-n^2 \eta(t)] \quad (48)$$

where $W(t) = \frac{T(t)-T_0}{\frac{Q}{\rho C_p L}} = \frac{T(t)-T_0}{T_{max}}$ and $\eta(t) = \frac{\pi^2 \alpha t}{L^2}$ are dimensionless parameters, $T_{max}$ is the maximum temperature at the rear side. The thermal diffusivity can be determined for a known specimen thickness $L$ and the time $t_{1/2}$ at which the temperature reaches half the maximum value:

$$\alpha = \frac{1.38 L^2}{\pi^2 t_{1/2}} \quad (49)$$

The laser flash method relies on a standard test method E1461 [30] (which outlines the procedure for determining thermal diffusivity) and standard ASTM practice E2585 [31] (which provides practical guidance to complement ASTM E1461, including recommendations for data analysis, test setup optimization, and uncertainty evaluation). This technique was further refined using different time parameters, which is called laser flash method. The following equations are introduced to estimate the in-plane diffusivity [32], [33]:

$$\alpha = \frac{L^2}{t_{5/6}} \left[ 0.818 - 1.708 \frac{t_{1/3}}{t_{5/6}} + 0.885 \left(\frac{t_{1/3}}{t_{5/6}}\right)^2 \right] \quad (50)$$

$$\alpha = \frac{L^2}{t_{5/6}} \left[ 0.954 - 1.581 \frac{t_{1/2}}{t_{5/6}} + 0.558 \left(\frac{t_{1/2}}{t_{5/6}}\right)^2 \right] \quad (51)$$

$$\alpha = \frac{L^2}{t_{5/6}} \left[ 1.131 - 1.222 \frac{t_{2/3}}{t_{5/6}} \right] \quad (52)$$

where the thermal diffusivity $\alpha$ [m$^2$/s] corresponds to the mean value of the above three results; $L$ is the thickness of the sample; $t_{1/2}$, $t_{1/3}$, $t_{2/3}$, and $t_{5/6}$ correspond to times when temperature equals 1/2, 1/3, 2/3, and 5/6 of its maximum value, respectively. Although the Parker method is based on one-dimensional heat conduction model, it still maintains high measurement accuracy, as validated in numerous studies [34], [35], [36], [37].

### B. Dimensionless Process for Thermal Diffusivity Measurement Using Rectangular Pulse Excitation

The commonly-used method for thermal diffusivity measurement is to fit the acquired experimental results based on the analytical solution. The mainstream modalities include time-domain and frequency-domain fitting. Therefore, Parker's solution and Mandelis's solution (in the Supplement and [29]) play an important role. It should be noted that in this paper a proper analytical solution is developed considering convection and heat extraction process for a Dirac pulse, Eq. (41), and for an instantaneous step-function excitation, Eqs. (42)-(43), in the time-domain.

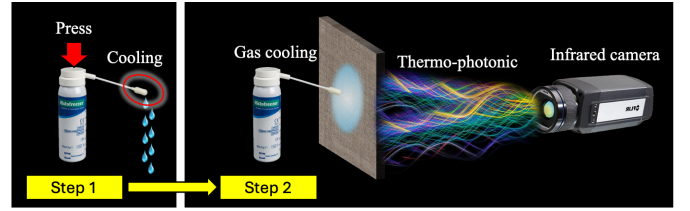

**Fig. 6.** Schematic of evaporative cooling method.

The schematic image of this arrangement is shown in Fig. 6. The evaporative cryocooling method involves two steps: Pressing an actuator to release nitrogen gas or a liquid; and bringing the cooling nozzle into contact with the sample surface to cool down the surface at the liquid freezing temperature of -55 °C. Simultaneously, the temperature variation on the rear side is record by an infrared camera. This approach ensures a more uniform cooling. Unlike the conventional laser flash method, the evaporative cryocooling approach induces a temperature decrease, thereby involving a reverse thermal behavior. The evaporative cryocooling experiment is consistent with Parker's solution (in the next section). During the short nozzle-sample contact, the very high heat-transfer coefficient at the surface drives a rapid ("instantaneous" on the time scale) temperature drop.

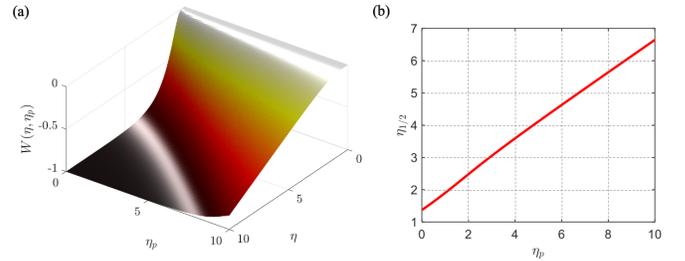

**Fig. 7.** Proposed thermal diffusivity measurement method: (a) The relationship between normalization amplitude and $(\eta, \eta_p)$; (b) The relationship between $\eta$ and $\eta_p$ at $|W|_{max}/2$. $W$ is dimensionless temperature, $(\eta, \eta_p)$ is dimensionless variables.

The laser flash method and Parker's method use flash lamps/impulse laser to measure thermal diffusivity based on the analytical solution in Eq. (7). Here, we extend these methods to instantaneous step-function excitation under the thermal heating or cooling mode. For the step-function response, Eq. (11) can be reduced to:

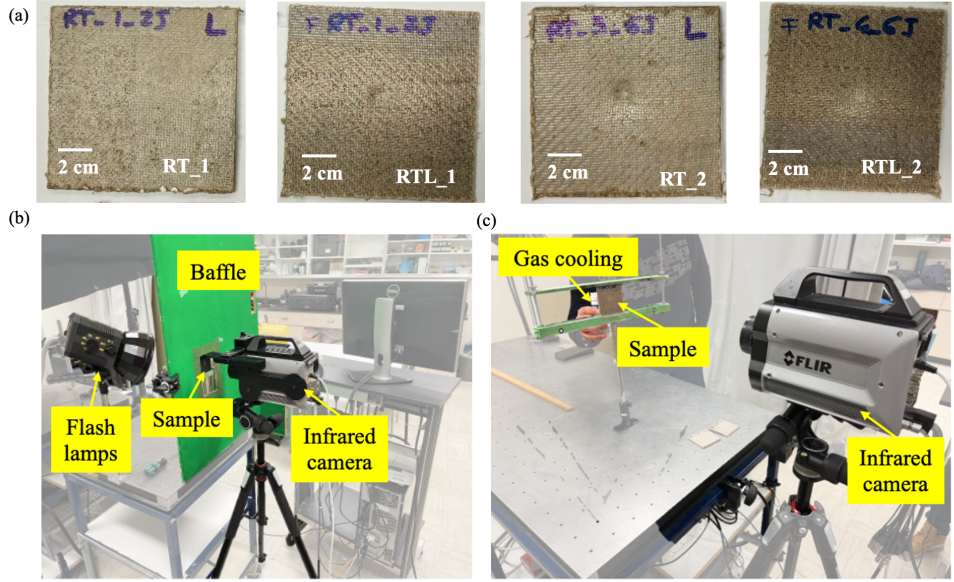

**Fig. 8.** Experimental configurations and simulations for thermal diffusivity measurements: (a) Samples. (b) Experimental setup for laser flash method. (c) Experimental setup for evaporative cryocooling method.

$$W(\eta,\eta_p) = \begin{cases} \frac{\eta}{\eta_p} + 2\sum_{n=1}^{\infty} \frac{(-1)^n}{n^2\eta_p}(1 - e^{-n^2\eta}), & \eta \leq \eta_p \\ 1 + 2\sum_{n=1}^{\infty} \frac{(-1)^n}{n^2\eta_p}(e^{-n^2(\eta-\eta_p)} - e^{-n^2\eta}), & \eta > \eta_p \end{cases} \quad (53)$$

where the dimensionless parameter $W(\eta,\eta_p)$ is defined as: $W(\eta,\eta_p) = \frac{(T(L,t)-T_0)\tau_p}{T_{max}}$, $T_{max} = \frac{Q_0\tau_p}{\rho c L}$, $\eta(t) = \frac{\pi^2 \alpha t}{L^2}$, and $\eta_p(t) = \frac{\pi^2 \alpha \tau_p}{L^2}$. For step-function cooling excitation, $W(\eta,\eta_p)$ can be calculated as follows:

$$W(\eta,\eta_p) = \begin{cases} -\frac{\eta}{\eta_p} - 2\sum_{n=1}^{\infty} \frac{(-1)^n}{n^2\eta_p}(1 - e^{-n^2\eta}), & \eta \leq \eta_p \\ -1 - 2\sum_{n=1}^{\infty} \frac{(-1)^n}{n^2\eta_p}(e^{-n^2(\eta-\eta_p)} - e^{-n^2\eta}), & \eta > \eta_p \end{cases} \quad (54)$$

It is possible to find the two variables $(\eta,\eta_p)$ in Eqs. (33) and Eq. (34) if the sample thickness is known. This is the difference between Parker's method and the step-function method. The latter requires foreknowledge of the heating time (rectangular pulse width). According to the solution in Eq. (54), a 2D thermal map for different heating times can be calculated as shown in Fig. 7(a). Next, simulating Eq. (54), yields the relationship between excitation time $\eta_p$ a feature time $\eta_{1/2}$ (half-maximum time). It should be noted that when the excitation time $\eta_p \to 0$, the half-maximum time equals 1.38. This fact reveals that Parker's solution is just a particular limiting case of our proposed solution, Eq. (11), when $\eta_p \to 0$. Furthermore, it is clear from the foregoing discussion that the excitation time $\eta_p$ must be known when using a rectangular pulse to measure the thermal diffusivity. This is so because there are two variables in Eqs. (53) and (54), and the half-maximum time $\eta_{1/2}$ increases approximately linearly with excitation time $\eta_p$.

In summary, it has been shown that the widely used Parker's half-maximum time equation represents a limiting case of Eqs. (53) and (54) which are solutions to the BVP of Eqs. (9)-(11) in the model developed in sections 2.2 above. The derivation of dimensionless process for heat conduction model considering convection is shown in Supplement.

## IV. EXPERIMENTAL SETUPS

### A. Samples

Four composite plates, polylactic acid/polybutylene adipate-co-terephthalate (PLA/PBAT)-based materials, were tested in this study: RT_1, RT_2, RTL_1, and RTL_2, as shown in Fig. 8(a). RT_1 and RTL_1 were subjected to 2 J impact energy, while RT_2 and RTL_2 experienced 6 J impact energy. The main distinction between these samples lies in the aging process: RT_1 and RT_2 were aged for one month in a salt spray chamber at room temperature, whereas RTL_1 and RTL_2 were unaged. All samples were composed of a PLA80%-PBAT20% matrix reinforced with flax fibers. For the polymer film production, the A500 matrix (PLA/PBAT 80/20 loaded with 10 wt% $CaCO_3$) was pre-dried at 70 °C for four hours. Films were extruded using a thermal profile along the screw, with temperatures set at 165 °C, 175 °C, 180 °C, 170 °C, and 165 °C (from the hopper to the die head), and a screw speed of 80 rpm. The resulting films had a thickness of approximately 100 μm.

To fabricate the laminates, samples were prepared by alternately stacking pre-dried A500 films (pre-dried at 70 °C for four hours) and flax fiber reinforcements (200 g/m²). The assembly, consisting of eight layers of natural fibers, was consolidated under the following conditions: processing temperature of 180 °C, and a stepwise pressure profile of 1, 5, 10, 15, 20, 25, and 30 bar (each maintained for 2 minutes). Final cooling was performed at ambient temperature under 40 bar pressure.

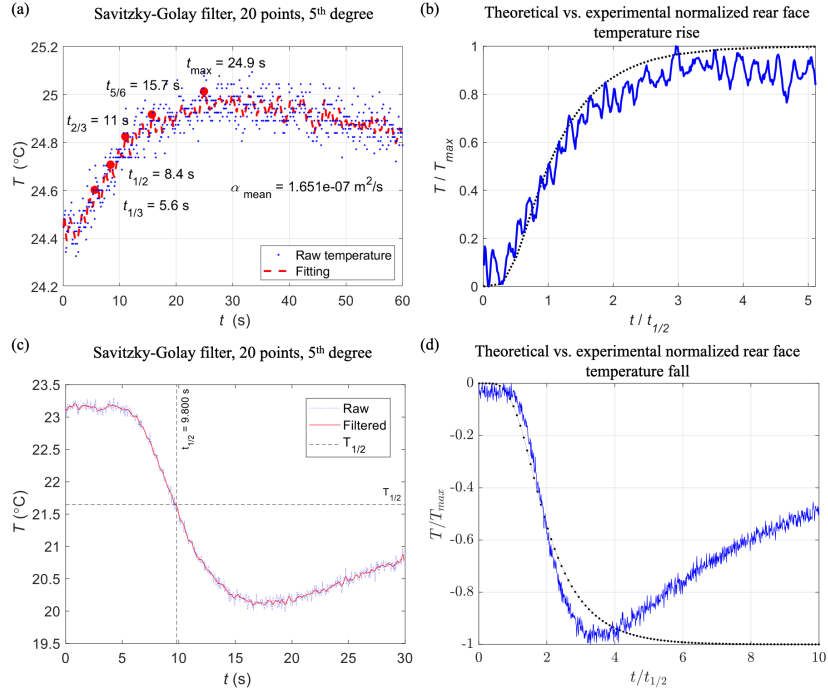

**Fig. 9.** Thermal diffusivity measurement results for RT_1: (a) Temperature profile fitting based on Savitzky-Golay filter (Parker's method); (b) The comparison between theoretical and experimental results (Parker's method); (c) Temperature profile fitting based on Savitzky-Golay filter (evaporative cryocooling method); (d) The comparison between theoretical and experimental results (evaporative cryocooling method).

TABLE II
Thermal diffusivity measurement based on Parker's method and evaporative cryocooling method.

| Materials | RT_1 | | RT_2 | | RTL_1 | | RTL_2 | |
|---|---|---|---|---|---|---|---|---|
| Damage | Damage | Sound | Damage | Sound | Damage | Sound | Damage | Sound |
| Parker's method | $1.50\times10^{-7}$ | $1.65\times10^{-7}$ | $1.75\times10^{-7}$ | $1.68\times10^{-7}$ | $1.35\times10^{-7}$ | $1.51\times10^{-7}$ | $1.42\times10^{-7}$ | $1.58\times10^{-7}$ |
| Proposed method | $1.67\times10^{-7}$ | $2.06\times10^{-7}$ | $2.56\times10^{-7}$ | $2.50\times10^{-7}$ | $1.52\times10^{-7}$ | $1.95\times10^{-7}$ | $1.70\times10^{-7}$ | $2.09\times10^{-7}$ |
| Absolute error | $0.17\times10^{-7}$ | $0.41\times10^{-7}$ | $0.71\times10^{-7}$ | $0.82\times10^{-7}$ | $0.17\times10^{-7}$ | $0.44\times10^{-7}$ | $0.28\times10^{-7}$ | $0.51\times10^{-7}$ |

*B. Experimental Setups*

Two experimental methods were designed in this work, including the photothermal radiometry technique and the evaporative cryocooling method, as shown in Fig. 8(b) and (c). A cooled infrared camera (FLIR X8501sc, 3-5 μm, InSb, NEdT < 20 mK, 1280 × 1024 pixels) for both laser flash and evaporative cryocooling experiments was used. Two Xenon flash lamps (Balcar, 6.4 kJ for each, 2 ms) have been used for photothermal experiments.

V. RESULTS AND DISCUSSION

*A. Parker's Methods for Thermal Excitation*

The laser flash method (or Parker's method) is the commonly used modality for thermal diffusivity measurements. In this study, it was also used as a reference for comparison, rather than as an absolute standard, to evaluate the performance of different methods. Temperature profiles were always accompanied by ambient and infrared camera noise. For this work a Savitzky-Golay denoising filter [38], [39] was employed as shown in Fig. 9(a). This filter smooths the data by fitting successive subsets of adjacent points with a low-degree polynomial, preserving important features such as peak height and width while reducing high-frequency noise. Different half-maximum time values were extracted to calculate the thermal diffusivity based on Eqs. (30)-(32). It should be noted that the $\alpha_{mean}$ in Fig. 8(a) denotes the average value of results from Eqs. (30)-(32) which when substituted in Eq, (29) yields the thermal diffusivity value $\alpha_{mean} = 1.651 \times 10^{-7}$ m²/s. This theoretical result is in good agreement with the experimental result, as shown in Fig. 9(b).

*B. Proposed Evaporative Cryocooling Methods*

The evaporative cryocooling method mentioned in Section 3.2 was used to measure thermal diffusivity. The goal was to check the agreement and/or accuracy of this method against Parker's method. From the required equipment and derived formalisms, the proposed (evaporative cryocooling) method exhibits better feasibility and physical reliability comparing with conventional Parker's method. The half-maximum time was extracted by measuring the location of $T_0 + \frac{1}{2}(T_{min} - T_0)$, as shown in Fig. 9(c). The feature coefficient has been calculated based on Eq. (34) and is demonstrated in Fig. 7(b). Here, the evaporative cryocooling time is approximately 1 s, and the corresponding coefficient $\eta_{1/2}$ is 1.8917 which yields the thermal diffusivity value $\alpha = 1.8351 \times 10^{-7}$ m²/s. It is clear that the result calculated by evaporative cryocooling is similar to that calculated by Parker's method within only 11.15% relative error. This validates the accuracy of the proposed method. Then, the theoretical solution was compared with the experimental result, as shown in Fig. 9(d) and it was found that the theoretical calculation was in good agreement with the experimental result (> 80% R-square before the minimum temperature value). It should be noted that the theoretical curves were not fitting based on experimental results. Instead, the therotical curves were calculated by substituting the given theoretical thermal diffusivity and thickness into Eqs. (9)-(11).

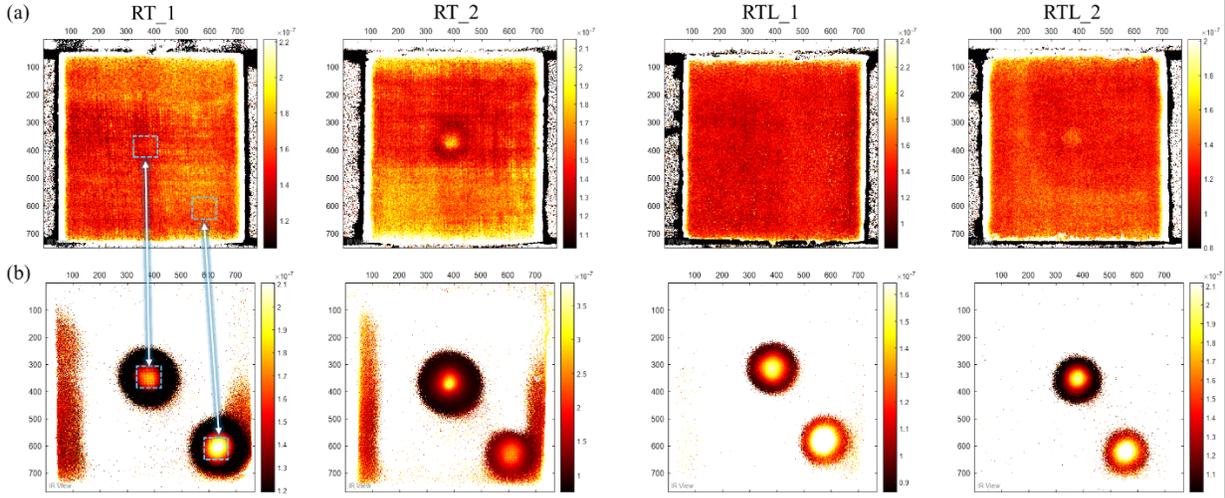

**Fig. 10.** Thermal diffusivity measurement results for RT_1, RT_2, RTL_1, and RTL_2: (a).

*C. Quantitative Comparison*

The above signal processing is for single pixel. It can be readily extended to all pixels. The overall thermal diffusivity maps are shown in Fig. 10. The mean filtering was employed to remove the noise from outliers in Fig. 10. According to the thermal diffusivity map, the difference caused by the aging process and impact energy is obvious: The aging process increases the thermal diffusivity anisotropy of composite materials. Furthermore, the effects of fiber orientation in RT_1 and RT_2 are clearly observable. It can be concluded that the aging process reduces the resistance to impact damage of composite materials. The thermal diffusivity measurement results from evaporative cryocooling method are shown in Fig. 10(b). Two regions were excited simultaneously, including those with and without impact damage. Unlike Parker's method, there is no outlier caused by thermal noise. Next, a small window from each result in Fig. 10(a) and (b) was selected to compare Parker's and the present results. The average values of RT_1, RT_2, RTL_1, and RTL_2 are shown in Table 2. According to results from Parker's method and evaporative cryocooling method, it is clear that the proposed method has high feasibility and accuracy. The maximum of absolute error is no more than 0.82×10⁻⁷. Of note, the results calculated by the Parker's method were processed by mean filtering. The Parker's method significantly relies on the temperature value at feature time. Even if the temperature value at this feature time has weak noise, the final results will deviate significantly. However, the temperature profile of the proposed method is relatively smooth. Because the heating area is focused on a small area and diffuse laterally.

V. CONCLUSION

In this study, we presented a comprehensive theoretical analysis for thermal response under thermal/cooling Dirac pulse and rectangular pulse excitations. The study revealed a hidden assumption made in Parker's theory shown to be that of a Dirac delta function impulsive boundary condition. Furthermore, it was found that the classic heat conduction model (Parker's solution) was in good agreement with a more rigorous formalism modelled by eigenfunction-eigenvalue based solutions under physically well-posed boundary conditions, especially before the "peak point" of the temperature profile curves. An explicit solution was derived using a dimensionless approach for rectangular pulse excitation and it was found that Parker's method is equivalent to the rectangular pulse method in the limit when the pulse duration tends to 0. Furthermore, a novel excitation approach was proposed which uses an evaporative cryocooling method. The results demonstrated high feasibility and accuracy in comparison with results from Parker's method where the relative error is no more than 11.15%.

SUPPLEMENTARY

*A. Heat Conduction Model Based on Green's Function Method*

To better describe a real thermal system under open boundaries, finite heat losses, and non-ideal excitations, we provide the frequency-domain analytical solution based on the Green's function method [34,35], as shown in Fig. S1.

Considering a one-dimensional solid under homogeneous boundary conditions of the third kind (see Fig. S1), in which a spatially impulsive, thermal-wave source oscillating at angular frequency ω is introduced at the location $x = x_0$, the Green-function derivation is described as follows:

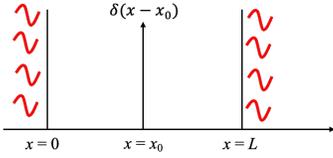

**Fig. S1.** Theoretical model using Green's function method. $\delta(x - x_0)$ denotes the impulse function at location $x_0$, $k$ is the thermal conductivity, $G$ is the Green's function, $\omega$ is the angular frequency, $h_1$ and $h_2$ denote the convective coefficients at left and right surfaces.

$$\begin{cases} \frac{d^2}{dx^2}G(x|x_0;\omega) - \sigma^2 G(x|x_0;\omega) = -\frac{1}{\alpha}\delta(x-x_0), \\ k\frac{d}{dx}G(x|x_0;\omega)|_{x=0} = h_1 G(0|x_0;\omega), \\ -k\frac{d}{dx}G(x|x_0;\omega)|_{x=L} = h_2 G(L|x_0;\omega) \end{cases} \quad (S1)$$

where $\sigma = (1+i)\sqrt{\omega/2\alpha}$ is the complex wavenumber, $h_1$ and $h_2$ are the convective coefficient of the front and rear surface, and $k$ is the thermal conductivity. Eq. (S1) can be solved directly [29]:

$$G(x|x_0;\omega) = \frac{1}{2\alpha\sigma}\left[\frac{e^{-\sigma|x-x_0|} + R_1 e^{-\sigma(x+x_0)} + R_1 R_2 e^{-\sigma[2L-|x-x_0|]} + R_2 e^{-\sigma[2L-(x+x_0)]}}{1 - R_1 R_2 e^{-2\sigma L}}\right], 0 \le x \le L \quad (S2)$$

where $R_1 \equiv \frac{k\sigma - h_1}{k\sigma + h_1}$ and $R_2 \equiv \frac{k\sigma - h_2}{k\sigma + h_2}$. Since the heat source is on the front surface, i.e., $x_0 = 0$, the Green's function can be reduced to:

$$G(x;\omega) = \frac{1}{2\alpha\sigma}\left[\frac{(1+R_1)(e^{-\sigma x} + R_2 e^{-\sigma(2L-x)})}{1 - R_1 R_2 e^{-2\sigma L}}\right], 0 \le x \le L \quad (S3)$$

For transmission mode, i.e., $x = L$, the resulting transient temperature distribution of a temporal impulse excitation can be calculated:

$$T^*(L,t) = \frac{1}{2\pi}\int_{-\infty}^{\infty} G(L;\omega)e^{i\omega t}d\omega$$
$$= \frac{1}{4\pi\alpha}\int_{-\infty}^{\infty}\frac{(1+R_1)(1+R_2)}{\sigma(1-R_1 R_2 e^{-2\sigma L})}e^{-\sigma L + i\omega t}d\omega \quad (S4)$$

The thermal-wave impulse response can be used to calculate any photothermal response to an arbitrary excitation waveform $E(t)$ through the convolution of $E(t)$ with $T^*$:

$$T_{arbitrary}(t) = E(t) * T^*(L,t) \quad (S5)$$

where $*$ represents convolution operator. In this study, the excitation signal is a rectangular pulse waveform:

$$E(t) = Q_0[H(t) - H(t-\tau_p)] \quad (S6)$$

The thermal response under rectangular pulse excitation can be given as:

$$T(L,t) = \frac{Q_0}{4\pi\alpha}\int_{-\infty}^{\infty}\Phi(\omega)\frac{e^{i\omega t}-1}{i\omega}d\omega, \ 0 \le t < \tau_p \quad (S7)$$

$$T(L,t) = \frac{Q_0}{4\pi\alpha}\int_{-\infty}^{\infty}\Phi(\omega)\frac{e^{i\omega t}-e^{i\omega(t-\tau_p)}}{i\omega}d\omega, \ t \ge \tau_p \quad (S8)$$

where $\Phi(\omega) = \frac{(1+R_1)(1+R_2)}{\sigma(1-R_1 R_2 e^{-2\sigma L})}e^{-\sigma L}$. The above equations can also be unified as:

$$T(L,t) = \frac{Q_0}{4\pi\alpha}\int_{-\infty}^{\infty}\Phi(\omega)\frac{e^{i\omega t}-e^{i\omega(t-\min(t,\tau_p))}}{i\omega}d\omega \quad (S9)$$

It should be noted that the above equations are based on $T_0 = 0$. For real thermal systems, the initial temperature $T_0$ should be added in Eq. (S7)-(S9).

The frequency-domain solution under Dirac pulse excitation is:

$$T^*(L,t) = -\frac{1}{4\pi\alpha}\int_{-\infty}^{\infty}\frac{(1+R_1)(1+R_2)}{\sigma(1-R_1 R_2 e^{-2\sigma L})}e^{-\sigma L + i\omega t}d\omega \quad (S10)$$

The frequency-domain solution under rectangular pulse excitation is:

$$T(L,t) = -\frac{Q_0}{4\pi\alpha}\int_{-\infty}^{\infty}\Phi(\omega)\frac{e^{i\omega t}-e^{i\omega(t-\min(t,\tau_p))}}{i\omega}d\omega \quad (S11)$$

*B. The Equivalent Between Parker's Assumption and Dirac Pulse Heat Injection*

Parker's assumption is:

$$f(x) = \begin{cases} \frac{Q_0}{\rho C_p g}, & 0 < x < g, \\ 0, & g < xL, \end{cases} \quad (S12)$$

As mentioned before, for the Neumann-Neumann problem, one finds for $n \ge 1$ (with $\phi_n(x) = \cos(\frac{n\pi x}{L})$ and $\int_0^L \phi_n^2 dx = L/2$):

$$A_n(0^+) = \frac{2}{\rho C_p L}Q_0 \quad (S13)$$

For the zero mode:

$$A_0(0^+) = \frac{Q_0}{\rho C_p L} \quad (S14)$$

Thus the solution reads:

$$T(x,t) = T_0 + A_0(0^+) + \sum_{n=1}^{\infty}A_n(0^+)e^{-\alpha\lambda_n^2 t}\cos(\frac{n\pi x}{L}) \quad (S15)$$

Projecting the initial profile $f(x)$ onto $\phi_n$ gives:

$$A_n(0) = \frac{1}{\int_0^L \phi_n^2(x)dx} = \frac{2}{L}\int_0^g \frac{Q_0}{\rho C_p g}\cos(\frac{n\pi x}{L})dx \quad (S16)$$

The Eq. (16) can be simplified as:

$$A_n(0) = \frac{2Q_0}{\rho C_p L}\frac{\sin(\frac{n\pi g}{L})}{\frac{n\pi g}{L}} \equiv \frac{2Q_0}{\rho C_p L}S_n(g) \quad (S17)$$

where $S_n(g) = \frac{\sin(\frac{n\pi g}{L})}{\frac{n\pi g}{L}}$. And for the zero mode:

$$A_0(0) = \frac{1}{L}\int_0^L f(x)dx = \frac{Q_0}{\rho C_p L} \quad (S18)$$

Now, we can find: 1) The equivalent of zero-mode:

$$A_0(0) = \frac{Q_0}{\rho C_p L} = A_0(0^+) \quad (S19)$$

2) For high-order modes:

$$\lim_{g \to 0} A_n(0) = \frac{2Q_0}{\rho C_p L} = A_n(0^+) \quad \text{for } n \ge 1 \quad (S20)$$

where $\lim_{g \to 0} S_n(g) = 1$. Therefore, every modal coefficient from the thin-layer initial condition tends exactly to the coefficient induced by the boundary flux pulse. The resulting series for $T(x,t)$ is therefore identical in that limit. We can conclude that by direct eigenfunction projection one sees that concentrating the same total heat $Q_0$ in a vanishingly thin layer near $x = 0$ yields, in the limit $g \to 0$, identical modal amplitudes to those produced by an instantaneous flux pulse at the boundary. Thus, the two solution methods are mathematically equivalent.

*C. Dimensionless Process for Heat Conduction Model Considering Convection*

The dimensionless equation of Eq. (45) becomes:

$$W(\eta) = -\sum_{n=1}^{\infty}\left[\frac{2[(\mu_n)^2+Bi^2]}{(\mu_n)^2+Bi^2+\mu_n \cdot Bi}\right]\cos(\mu_n)e^{-(\mu_n)^2\eta} \quad \text{(S21)}$$

where $\eta = \alpha t/L^2$ is the dimensionless time, $W = \frac{T-T_0}{q_0 L/k}$ is the dimensionless temperature, $Bi = hL/k$ is the Biot number, $\mu_n = \lambda_n L$ is the dimensionless eigenvalue, which can be calculated by $\tan(\mu_n) = \frac{2Bi\mu_n}{(\mu_n)^2 - Bi^2}$. For thermal response of rectangular pulse excitation in Eq. (46) and (47), the dimensionless equations become:

$$W(\eta,\eta_p) = \begin{cases} -\sum_{n=1}^{\infty}\left[\frac{2[(\mu_n)^2+Bi^2]}{(\mu_n)^2+Bi^2+Bi}\right]\cdot\frac{1-e^{-(\mu_n)^2\eta}}{(\mu_n)^2}\cos(\mu_n), & \eta \leq \eta_p \\ -\sum_{n=1}^{\infty}\left[\frac{2[(\mu_n)^2+Bi^2]}{(\mu_n)^2+Bi^2+Bi}\right]\cdot\frac{1-e^{-(\mu_n)^2\eta_p}}{(\mu_n)^2}e^{-(\mu_n)^2(\eta-\eta_p)}\cos(\mu_n), & \eta > \eta_p \end{cases}$$

$$\text{(S22)}$$

According to Eq. (S21) and (S22), it is possible to find that we cannot extract feature time since there is an unknown parameter, thermal conductivity, in $Bi = hL/k$.

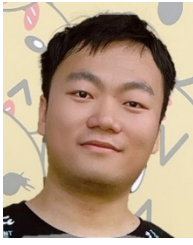
**Pengfei Zhu** (Graduate Student Member, IEEE) received the B.Eng. degree in engineering mechanics from North University of China, Taiyuan, China, in 2019, and the M.Eng. degree in solid mechanics from Ningbo University, Ningbo, China, in 2022. He is currently working toward the Ph.D. degree in electrical engineering with Université Laval, Québec, Canada.

His research interests include non-destructive testing, infrared thermography, deep learning, terahertz time-domain spectroscopy, and photothermal coherence tomography.

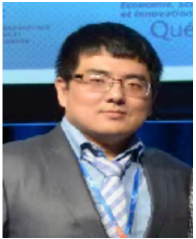
**Hai Zhang** (Member, IEEE) is a full professor at Harbin Institute of Technology, China. He received his BSc and MSc degrees from Shenyang University of Technology, China in 2004 and 2008, respectively, and his Ph.D. degree from Laval University, Canada, in 2017, where he is currently an adjunct professor. He was a Postdoctoral Research Fellow with University of Toronto, Canada. He was also a Visiting Researcher in Fraunhofer EZRT, Fraunhofer IZFP and Technical University of Munich, Germany.

His research interests include nondestructive testing, industrial inspection, medical imaging, infrared and terahertz spectroscopy, etc. He has authored or coauthored more than 150 technical papers in peer-reviewed journals and international conferences. He is also an Associate Editor for Infrared Physics and Technology, Measurement, and Quantitative InfraRed Thermography Journal, etc.

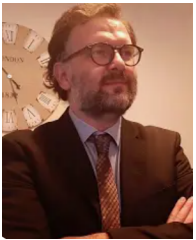
**Stefano Sfarra** received the Ph.D. degree in mechanical, management, and energy engineering from the University of L'Aquila (UNIVAQ), L'Aquila, Italy, in 2011. He worked as a Post-Doctoral Fellow at UNIVAQ until October 2017, where he became a Researcher in a fixed-term contract at the Department of Industrial and Information Engineering and Economics (DIIIE), UNIVAQ. Currently, he is an Associate Professor at DIIIE-UNIVAQ and an Adjunct Professor at Université Laval, Quebec, QC, Canada.

He is specialized in infrared thermography, heat transfer, speckle metrology, holographic interferometry, near-infrared reflectography, energy saving, and finite element simulation techniques. Concerning these research topics, he has published more than 200 papers in journals and international conferences.

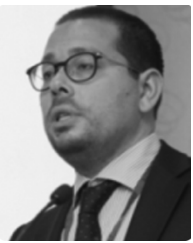
**Fabrizio Sarasini** received the Ph.D. degree in materials engineering from Sapienza University of Rome, Rome, Italy, in 2007. Since June 2016, he has been an Assistant Professor with the Faculty of Civil and Industrial Engineering, Sapienza University of Rome. He has more than 90 international peer-reviewed journal papers. His research interests include the impact response of composite materials for structural applications, the fiber/matrix interface adhesion, and the combination of natural fibers in hybrid composites.

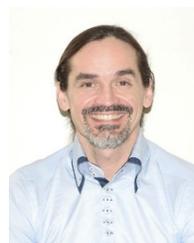
**Clemente Ibarra-Castanedo** received the M.Sc. degree in mechanical engineering (heat transfer) in 2000 from Université Laval, Quebec City, Canada, and the Ph.D. degree in electrical engineering (infrared thermography) in 2005 from the same institution. He is a professional researcher in the Computer Vision and Systems Laboratory at Université Laval and a member of the multipolar infrared vision Canada Research Chair (MIVIM). He has contributed to several research projects and publications in the field of infrared vision. His research interests are in signal processing and image analysis for the nondestructive characterization of materials by active/passive thermography, as well as near and short-wave infrared reflectography/transmittography imaging.

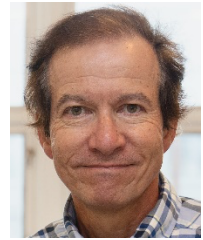
**Xavier Maldague** (Senior Member, IEEE) P.Eng., Ph.D. is full professor at the Department of Electrical and Computing Engineering, Université Laval, Québec City, Canada. He has trained over 50 graduate students (M.Sc. and Ph.D.) and contributed to over 400 publications. His research interests are in infrared thermography, NonDestructive Evaluation (NDE) techniques and vision / digital systems for industrial inspection. He is an honorary fellow of the Indian Society of Nondestructive Testing, fellow of the Canadian Engineering Institute, Canadian Institute for NonDestructive Evaluation, American Society of NonDestructive Testing. In 2019 he was bestowed a Doctor Honoris Causa in Infrared Thermography from University of Antwerp (Belguim).

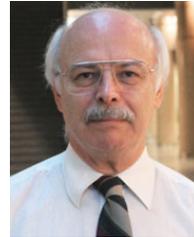
**Andreas Mandelis** FRSC, FCAE, FAPS, FSPIE, FAAAS, FASME, DF-IETI, PhD, is a full professor of Mechanical and Industrial Engineering; Electrical and Computer Engineering; and the Institute of Biomaterials and Biomedical Engineering, University of Toronto, and director of the Center for Advanced Diffusion-Wave and Photoacoustic Technologies at the University of Toronto. He is also the director of the Institute for Advanced Non-Destructive and Non-Invasive Technologies of the University. He has published more than 490 scientific papers in refereed journals and 190 proceedings papers in the fields of diffusion waves and photoacoustics. He has received numerous national and international prizes and awards and has several instrumentation and measurement methodology patents in photothermics, non-destructive evaluation, thermophotonics, optoelectronics, biophotoacoustics and new imaging methodologies.